# Modified ResNet Model for MSI and MSS Classification of Gastrointestinal Cancer

CH Sai Venkatesh[*], Caleb Meriga, M.G.V.L Geethika, T Lakshmi Gayatri, V.B.K.L Aruna[#]

Department of Electronics and Communication Engineering, VR Siddhartha Engineering College (Autonomous), JNTU Kakinada, India.

[*]178w1a0411@vrsec.ac.in, [#]aruna@vrsiddhartha.ac.in

*Abstract - In* **this work, a modified ResNet model is proposed for the classification of Microsatellite instability(MSI) and Microsatellite stability(MSS) of gastrointestinal cancer. The performance of this model is analyzed and compared with existing models. The proposed model surpassed the existing models with an accuracy of 0.8981 and F1 score of 0.9178.**

***Keywords-ResNet, MSI, MSS, Gastrointestinal cancer.***

## I. Introduction

During the "Genesis of Cancer" the word "Cancer" was rarely heard and we never thought that we would be hearing it so often. As per IARC(International Agency for Research on Cancer) 1 in 5 people develop cancer. Among all cancer related deaths Gastrointestinal Cancer constitutes to 35% of global cancer related deaths. Computer Vision was used to detect cancer tumors through histological images which drastically cut down both the time and money to carry out conventional testing methods[1-6].

Microsatellite is defined as the rudimentary repetitive sequence of the Deoxyribonucleic acid (DNA). DNA comprises of many microsatellites. DNA Mismatch Repair (MMR) is a system which monitors the replication process of microsatellites and DNA, if it finds any error in the DNA recombination and replication it performs repair with the help of MMR proteins. Failure of MMR leads to unstable microsatellites/DNA which is the genesis of cancer. Based on global genomic status cancer tumor is classified into 'Microsatellite instable' (MSI ) and 'Microsatellite Stable' (MSS) tumor. High amount of instability in tumor classifies it as MSI-H and it can be inherited, in which the immune cells are shut off from fully doing their job. By using 'Immunotherapy' MSI-H can be cured .In MSS the DNA in tumor cell has the same number of microsatellites that of a healthy cell, this can be cured by 'radiation' and 'chemotherapy'- treatments which are opposite to immunotherapy. 26.4% of gastrointestinal cancer patients are classified as MSI-H and the rest i.e. 73.6% as MSS .Therefore, detection of MSI or MSS of cancer has the same significance as detection of cancer to give appropriate treatment.

In this paper, we trained different pre-trained models and proposed a 'Modified ResNet' model to classify MSI or MSS. 192,000 histological images have been sorted into 10% for test, 80% for training and 10% for validation by 'Justin lin'[7]. The original data set is provided by "Kather, Jakob Nikolas" [8].

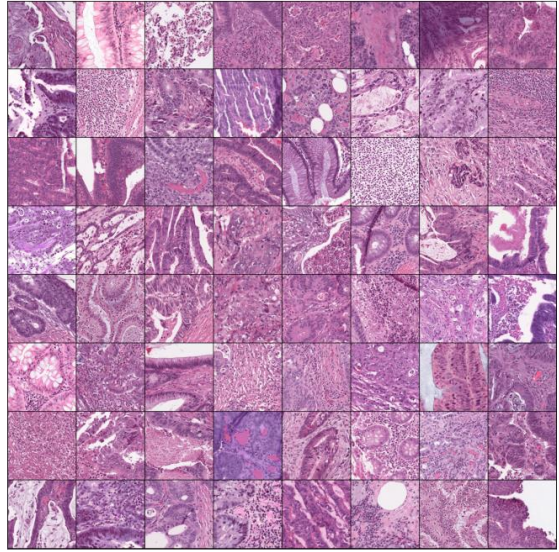

Fig.1 Sample images of MSI and MSS

## II. Model Description

To build a custom model, the performances of existing architectures on this data set have to be analyzed. It has been proven in many recent publications that 'ResNet' architecture is the cream of image classification[9]. In this paper two approaches are used, each containing a group of architectures to determine their performance on the current data set.

### A. Baseline Approach:

In this approach three baseline architectures were simulated they are 'Logistic Regression','4-Layer Feed Forward Neural Network' and 'Convolution Neural Network' (CNN). CNN architecture consists of five convolution layers, two linear layers with dropout rate of 0.5. Each convolution layer is followed by batch normalization. Since 'RELU' is computationally efficient, it is used as the activation function except for the output layer. 'Max pool 2D' with Stride two is used to reduce the image size by half.

B. *Transfer Learning Approach:*

VGG16 and various versions of ResNet pre-trained models are simulated in this approach. VGG16 architecture is simple and efficient which performed well in many computer vision Tasks[10]. ResNet and its various versions such as ResNet-18,34,50,101,152 are tested to find the sweet spot for this particular dataset. ResNet - due to identity mapping function there is no loss of the input image unlike VGG architecture the information is lost as it propagates deeper. 'RELU' activation and 'MaxPool 2D' with stride two is used in every model.

C. *Modified ResNet model:*

In this modified ResNet architecture a 2D convolution layer, maxpool 2D layer are followed by four residual blocks which are sequentially connected. An adaptive average pool 2D is used to convert into single dimension; a sigmoid function is used for binary classification. To ramp up the training, after every convolution layer batch normalization is used. For the first residual block stride is one and for the remaining residual blocks Stride two is taken. A convolution layer of 1*1 is employed to match input and output, when the sizes of input and output blocks are unequal. In fully connected network dropout with different dropout rates have been incorporated to reduce over fitting. The total architecture consists of 41 layers. Table I provides the parameters of the architecture.

Table I Modified ResNet Architecture

| Layer name | Specific's | Output Size |
|---|---|---|
| Convolution | 2D , (3*3,64,stride-2) | 112*112*64 |
| Maxpool | 2*2 ,Stride-2 | 56*56*64 |
| ResNet Block 1 | $\begin{bmatrix} 1*1 & 64 \\ 3*3 & 64 \\ 1*1 & 256 \end{bmatrix} *2$ | 56*56*256 |
| ResNet Block 2 | $\begin{bmatrix} 1*1 & 128 \\ 3*3 & 128 \\ 1*1 & 512 \end{bmatrix} *3$ | 28*28*512 |
| ResNet Block 3 | $\begin{bmatrix} 1*1 & 256 \\ 3*3 & 256 \\ 1*1 & 1024 \end{bmatrix} *5$ | 14*14*1024 |
| ResNet Block 4 | $\begin{bmatrix} 1*1 & 512 \\ 3*3 & 512 \\ 1*1 & 2048 \end{bmatrix} *2$ | 7*7*2048 |
| Average pool | Adaptive 2D (output size =(1,1)) | 1*1*2048 |
| FC1 | 2048 | |
| FC2 | 512 | |
| FC3 | 128 | |
| FC4 | 1,Sigmoid | |

# III. Results and Discussion

The provided dataset images are preprocessed, and are available in standard image size of 224*224*3. The data set is imported into 'Google Colaboratory' from 'Kaggle'. Code is written for every model using 'PyTorch' framework to obtain accuracy, F1 score and confusion matrix.

## A. Baseline Models:

Fig.2(a) shows the loss Vs number of epochs for logistic regression model, it can be seen that the training loss and validation loss remain constant at 39.0352 and 36.6357(at $10^{th}$ epoch). Fig.2(b) illustrates the loss Vs number of epochs for feed forward neural network model, it can be seen that as the number of epochs increase the training loss decreases but whereas the validation loss is fluctuating. The training loss and validation loss are 0.6473 and 0.6728(at $10^{th}$ epoch).The loss Vs number of epochs graph for CNN is shown in Fig.2(c) in this graph as we can see the at $10^{th}$ epoch both losses seem to be declining but when we increase the epochs it can be seen that they diverge which indicates over fitting shown in Fig.2(d) . The training loss and validation loss are 0.0613 and 0.6774(at $100^{th}$ epoch).

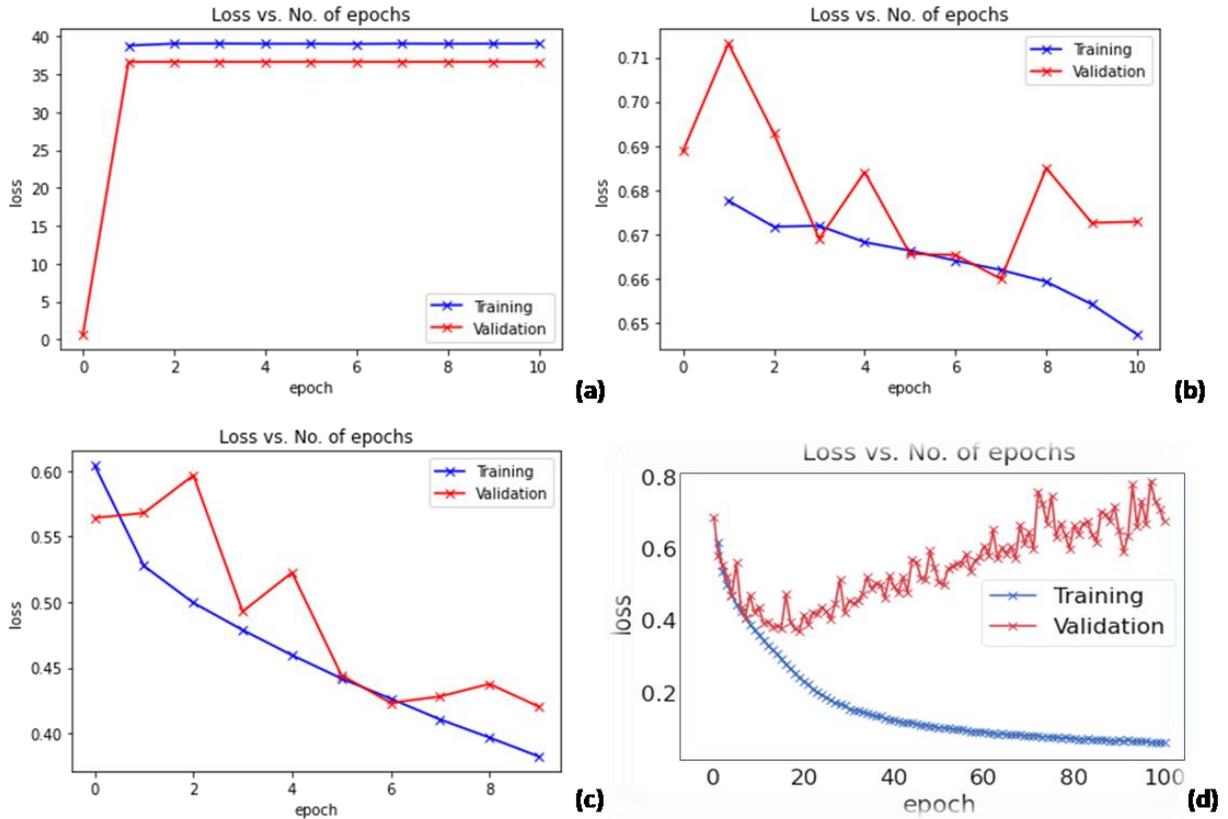

Fig.2. Loss Vs Number of epochs for Baseline models : (a)Logistic regression (b)Feed forward neural network (c)Convolution Neural Network upto 10 epochs (d) Convolution Neural Network upto 100 epochs

### B. Transfer learning Models:

VGG 16's loss vs number of epochs graph is shown in Fig.3, it can be observed that there is a spike at 16th epoch after that, the training loss and validation loss remain at 0.2829 and 0.2989(at 30th epoch).

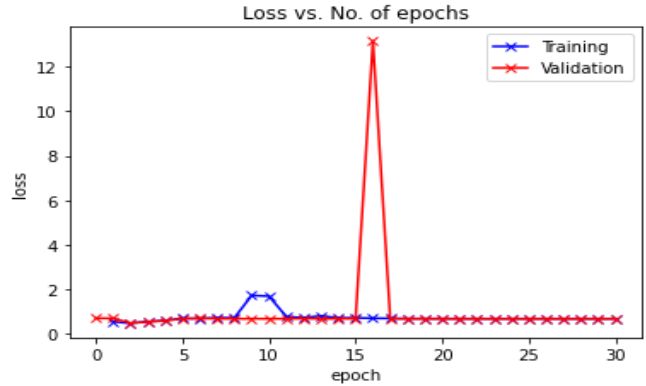

Fig.3. Loss Vs Number of epochs for VGG 16

Fig.4 contains the loss vs number of epochs graphs of Resnet-18,34,50,101,152. The training loss and validation loss of resnet18 are 0.2651 and 0.3111(at 40th epoch) shown in Fig.4(a) . The training loss and validation loss of resnet-34 are 0.3092 and 0.3203(at 20th epoch) shown in Fig.4(b) . Similarly for resnet50- 0.2829 and 0.2989(at 25th epoch) Fig.4(c), resnet101- 0.3102 and 0.3161(at 15th epoch) Fig.4(e), resnet152-0.3173 and 0.3307(at 15th epoch) Fig.4(d). From the graphs of the ResNet family it can be observed that as the order of the layers increase (such as 18,34,50...e.t.c.) the sooner the training and validation losses converge and be in proximity. In ResNet -18,34 base block is used and for the rest i.e.resnet-50,101,152 bottle neck block is incorporated.

### C. Modified ResNet Model:

From the results of the ResNet family, it is observed that the models having minimum number of layers have better performance even though they take more number of epochs to train. So with this in mind 'Modified ResNet Model' was built, initially base block was used but it didn't show any promise but when replaced by bottle neck block it surpassed the existing models. Here are a few parameters that were considered in this paper learning rate is 0.001,grad-clip =0.1,weight decay =1E-4,loss function = binary cross entropy and Adam optimizer. Fig.5 shows the loss Vs number of epochs for modified resnet model, it can be seen that the training loss and validation loss remain are less than other models 0.2149 and 0.2488(at 25th epoch).

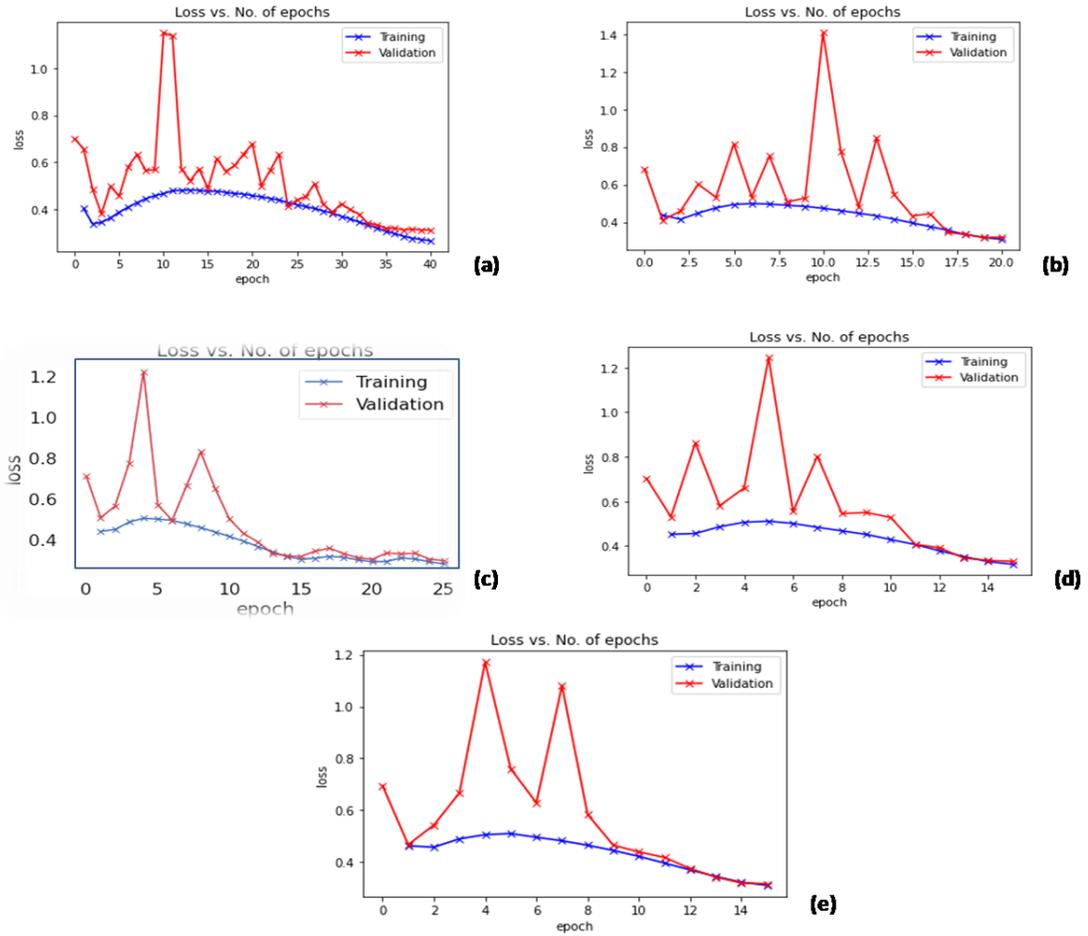

Fig.4. Loss Vs number of epochs of ResNet: (a)ResNet18 (b)ResNet34 (c)ResNet50 (d)ResNet152 (e) ResNet101

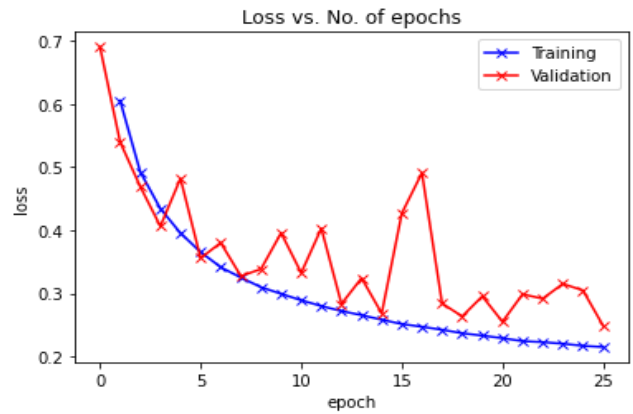

Fig.5. Loss Vs number of epochs of Modified ResNet Model

The confusion matrix is shown in Fig.6. Class-0 represents MSI and class-1 represents MSS. True Positive(TP) is defined as when MSI is predicted and the actual output is MSI. True Negative(TN) is defined as when MSS is predicted and the actual output is MSS. Similarly, when MSI is predicted but the actual output is MSS then it is False Positive(FP), when MSS is predicted but the actual output is MSI it's False Negative(FN) .

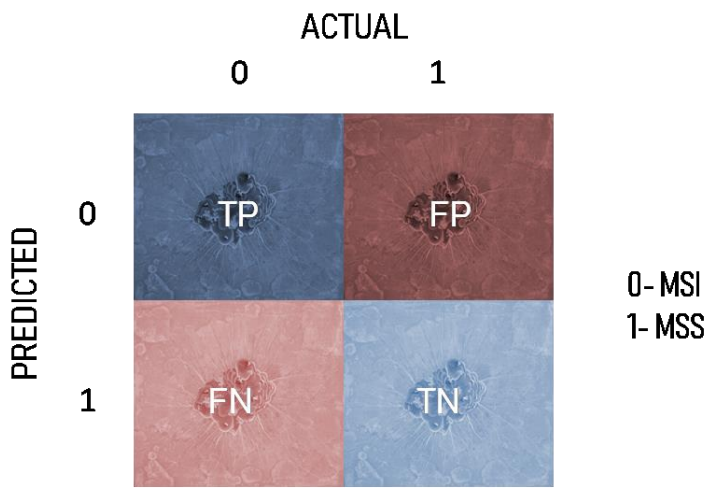

Fig.6. Confusion matrix

Table II  Confusion matrix values of simulated models

| Model | True Positive (TP) | False Positive (FP) | False Negative (FN) | True Negative (TN) |
|---|---|---|---|---|
| Logistic Regression | 0 | 7505 | 0 | 11728 |
| Feed Forward Neural Network | 523 | 6982 | 833 | 10895 |
| Convolution Neural Network | 5880 | 1625 | 1855 | 9873 |
| VGG16 | 0 | 7505 | 0 | 11728 |
| ResNet 18 | 6182 | 1323 | 1072 | 10656 |
| ResNet 34 | 6015 | 1490 | 1218 | 10510 |
| ResNet 50 | 6164 | 1341 | 972 | 10756 |
| ResNet 101 | 5940 | 1565 | 950 | 10778 |
| ResNet 152 | 5828 | 1677 | 943 | 10785 |
| Modified ResNet (proposed Model) | **6338** | 1167 | 792 | **10936** |

From table II it can be observed that the TP and TN values of the modified ResNet surpass other models which are desirable.

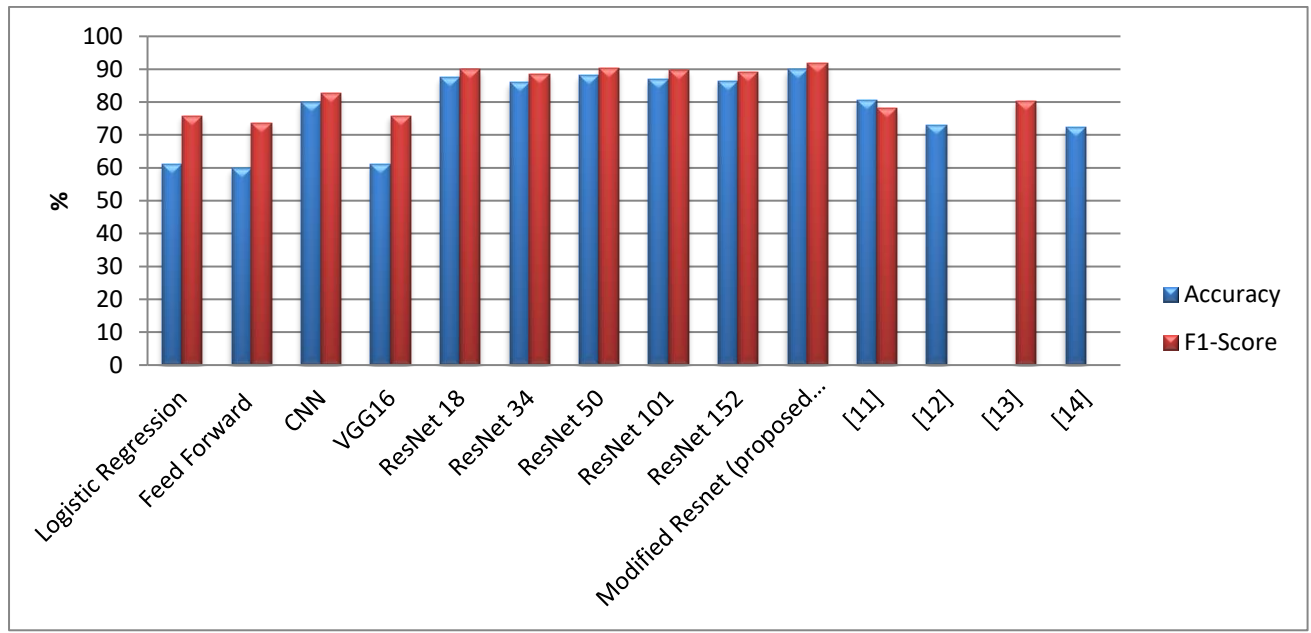

Fig.7. Comparison of accuracy and F1score with various simulated models and existing literature.

Fig.7. shows the comparison of accuracy and F1-score with different baseline, transfer learning models and existing literature. It is noted that the accuracy and F1-score of the modified ResNet model are 89.81% and 91.78% are better than remaining models.

## IV. Conclusion

A "Modified ResNet" model is proposed to classify MSI and MSS of Gastrointestinal cancer. The results were compared with baseline, transfer learning models and with existing literature. From the observations the proposed model exceeded other models with accuracy,F1-score,TP and TN of 89.81%,91.78% , 6338 and 10936. This model can be further improved by using full pre-activation ResNet blocks and data augmentation techniques. But this model cannot be used for dynamic inputs such as 'ECG'. This network can be implemented to classify MSI and MSS of other cancers.